%Paper: hep-th/9507009
%From: jrusso@vxcern.cern.ch
%Date: Mon, 3 Jul 1995 19:30:58 +0200
%Date (revised): Mon, 3 Jul 1995 20:17:46 +0200
%Date (revised): Wed, 12 Jul 1995 16:13:27 +0200

\input harvmac

%%%%%%%%%%%%%%%%%%%%%%%%%%%%%%%%%%%%%%%%%%%%%%%%%%%%%%%%%%%%%%%
%The following lines are needed to insert the accompanying figures
%in the paper. If you do not have epsf, then comment out the line
% ``\input epsf'', and print the figures separately.
\input epsf
\ifx\epsfbox\UnDeFiNeD\message{(NO epsf.tex, FIGURES WILL BE
IGNORED)}
\def\figin#1{\vskip2in}% blank space instead
\else\message{(FIGURES WILL BE INCLUDED)}\def\figin#1{#1}\fi
\def\ifig#1#2#3{\xdef#1{fig.~\the\figno}
\goodbreak\topinsert\figin{\centerline{#3}}%
\smallskip\centerline{\vbox{\baselineskip12pt
\advance\hsize by -1truein\noindent{\bf Fig.~\the\figno:} #2}}
\bigskip\endinsert\global\advance\figno by1}
%%%%%%%%%%%%%%%%%%%%%%%%%%%%%%%%%%%%%%%%%%%%%%%%%%%%%%%%%%%%%%%%

\def \w{\omega}

\def \s {\sigma}

\def \p {\phi}
\def \ha {\half}
\def \ov {\over}

\def \b {\beta}
\def \lr { \lref}
\def\ep{\varepsilon}
\def\vp {\varphi}
\def\dd {\partial }

\def\l{\lambda}
\def\ll{\lambda ^2}

\def \k {\kappa}
\def \h {\hbar }
\def\n{\noindent}
\gdef \jnl#1, #2, #3, 1#4#5#6{ { #1~}{ #2} (1#4#5#6) #3}

\def\np {  Nucl. Phys. }
\def \pl { Phys. Lett. }

\def \prl { Phys. Rev. Lett. }
\def \pr  { Phys. Rev. }

\lr \beken {J.D. Bekenstein, \pr D7 (1973) 2333; D9 (1974) 3292.  }
\lr \hawk {S. Hawking, Commun. Math. Phys. 43 (1975) 199.  }
\lr \thoof { G. 't Hooft,  \np B256 (1985) 727.  }
\lr \hooft { G. 't Hooft, Physica Scripta T36 (1991) 247;
 {\it Dimensional reduction in quantum gravity},
Utrecht preprint THU-93/26, gr-qc/9310006. }
\lr \sussk {L. Susskind,  {\it The world as a hologram},
  preprint SU-ITP-94-33, hep-th/9409089. }
\lr \susugl {L. Susskind and J. Uglum, \pr D50 (1994) 2700.}
\lr \suss {L. Susskind,  {\it Some speculations about entropy in
string theory},    RU-93-44, hep-th/9309145;
J.G. Russo and L. Susskind, \np B 437 (1995) 611; A.~Sen,
{\it Extremal black holes and elementary string states},
  TIFR-TH-95-19, hep-th/9504147; A. Peet, {\it Entropy and supersymmetry of $D$
dimensional extremal electric black holes versus string states},
  PUPT-1548, hep-th/9506200.}
\lr \sredn {M. Srednicki, \prl 71 (1993) 666;
V. Frolov and I. Novikov, \pr D48 (1993) 4545;
D. Kabat and M.J. Strassler, \pl B329 (1994) 46;
C. Callan and F. Wilczeck; \pl B333 (1994) 55;
J.L. Barb\' on and R. Emparan,   PUPT-1529, EHU-FT 95/5,
hep-th/9502155.}
\lr \rst {J.G. Russo, L. Susskind and L. Thorlacius,
\pr D46 (1992) 3444; \pr D47 (1993) 533.  }
\lr \stu {L. Susskind, L. Thorlacius and J. Uglum,
\pr D48 (1993) 3743.  }
\lr \membr {K. Thorne, R. Price and D. MacDonald, {\it Black holes:
the membrane paradigm} (Yale Univ. Press, New Haven, CT, 1986). }
\lr \entropy {E. Keski-Vakkuri and S. Mathur, \pr D50 (1994) 917;
T. Fiola, J. Preskill, A. Strominger
and S. Trivedi, \pr D50 (1994) 3987;  R.C. Myers, \pr D50 (1994) 6412;
J.D. Hayward, DAMTP-R94-61, gr-qc/9412065.
}
\lr \rstf  {J.G. Russo, L. Susskind and L. Thorlacius, \pl B292 (1992) 13.}
%\lr \veil  {J.G. Russo, \pr D49 (1994) 5266.}
\lr\wald {R. M. Wald, {\it General Relativity} (University of
Chicago Press, Chicago, 1984).}
 \lr \svv {K. Schoutens, H. Verlinde and E. Verlinde, {\it
Black hole evaporation and quantum gravity},
CERN-TH.7142/94.}

\baselineskip8pt
\Title{\vbox
{\baselineskip 6pt{\hbox{CERN-TH/95-179}} {\hbox{hep-th/9507009}} {\hbox{
   }}} }
{\vbox{\centerline {Entropy and black hole horizons}
}}

\vskip -20 true pt

\centerline  { {J.G. Russo\footnote {$^*$} {e-mail address:
 jrusso@vxcern.cern.ch
} }}

 \smallskip \bigskip

\centerline{\it  Theory Division, CERN}
\smallskip

\centerline{\it  CH-1211  Geneva 23, Switzerland}

\bigskip\bigskip\bigskip
\centerline {\bf Abstract}
\bigskip

The standard approach of counting the number of eigenmodes of $N$ scalar
fields near the horizon is used as a basis to provide
a simple statistical mechanical derivation of the black hole entropy
in two and four dimensions.
The Bekenstein formula $S={A\ov 4G\h}$ and the two-dimensional entropy
$S=2M/\l\h $ are naturally obtained (up to a numerical constant of order 1).
This approach provides an explanation on why the black hole entropy is of
order $1/\h $ and why it is independent of the number of field-theoretical
degrees of freedom.

\medskip
\baselineskip8pt
\noindent

\Date {July 1995}

%\draftmode
\noblackbox
\baselineskip 14pt plus 2pt minus 2pt
%\baselineskip 20pt plus 2pt minus 2pt
%%%%%%%%%%%%%%%%%%%%%%%%%%%%%%%%%%%%%%%%%%%%%%%%%%%%%%%%%%%%%%%%

\vfill\eject
%%%%%%%%%%%%%%%%%%%%%%%%%%%%%%%
\newsec{Horizons of evaporating black holes}
%%%%%%%%%%%%%%%%%%%%%%%%%%%
An important problem in quantum gravity is
understanding the origin of Bekenstein entropy of black holes,
\eqn\uno{
S={A\ov 4 G \h}  \ ,\ \ \ A=4\pi r_s^2=16\pi G^2M^2 \ .
}
Bekenstein suggested that $S$, more than a mere analogy
 to the thermodynamical entropy, should dictate
the number of possible quantum states of a black hole \beken .
This interpretation was put on a firmer ground after Hawking's discovery of
black hole radiation \hawk .  Subsequently it was argued that
a statistical mechanical origin of the
Bekenstein entropy  requires an enormous reduction of the number of
gauge-invariant degrees of freedom in quantum gravity, to the extent that
one space dimension might be superfluous \refs {\hooft , \sussk }.
Recently different statistical mechanical derivations of the Bekenstein
entropy,
as a geometric or entanglement entropy, have been investigated
(see e.g. refs. \refs {\thoof, \susugl , \suss , \sredn}).
The present derivation is closer in  spirit to the ``brick wall" model
of ref. \thoof , but in addition we shall take into account the black hole
evaporation. Although an infinite result was found in
\refs {\thoof , \susugl}, there the black hole was considered
as a fixed background. Here it will be shown that the result is finite upon
incorporating the back reaction on the geometry due to Hawking radiation.
The standard entropy  formula of   two-dimensional dilaton gravity will
  emerge
(up to a numerical factor of order 1).
While the inclusion of  back-reaction effects will lead
to a finite result in four-dimensional gravity as well, the resulting
quantum-field-theory entanglement entropy
is much larger than the Bekenstein entropy.
We will argue that the correct Bekenstein entropy naturally follows from the
assumption that the quantum mechanical information is encoded in the
Hawking radiation.

Classically, in the Schwarzschild frame of an external observer,
infalling matter reaches the event horizon in an infinite time.
At quantum level, assuming that at some moment the black hole
completely disappears, matter crosses the event horizon at some
calculable, finite Schwarzschild time (this time can be explicitly
obtained, e.g. in the model of ref. \rst ).
Infinite Schwarzschild time now corresponds to another horizon,
which is just beyond the event horizon, and we shall
call it (following ref. \stu ) the {\it ultimate} horizon (see also refs.
\refs {\hawk, \rst} ).
In the classical theory the event horizon and the ultimate horizon
coincide.
When the evaporation of the black hole is taken into account,
quantities which diverged on the classical black hole geometry
 (due to the presence of the event horizon), such as redshifts,
are now finite.
In particular, the integral
of the outgoing energy density flux is no longer infinity, since the Hawking
flux measured by any time-like outside observer stops after he enters into the
vacuum region (in fact this integral reproduces, as expected, the  total
ADM energy  of original collapsing matter).  A natural question is
whether also the entropy associated
with quantum fields near the horizon is finite when the black hole
evaporation is incorporated. This turns out to be  the case.
However, in four dimensions the resulting field theoretic entropy will
still be far larger than the Bekenstein entropy, and a cut-off will
be needed.

%$$\delta_a s=\int _{r_0}^{r_0+a} dr \sqrt{g_{rr}}$$

\ifig\fone{Penrose diagram for the process of formation and evaporation of a
black hole.}
{\epsfxsize=5.0cm \epsfysize=7.1cm \epsfbox{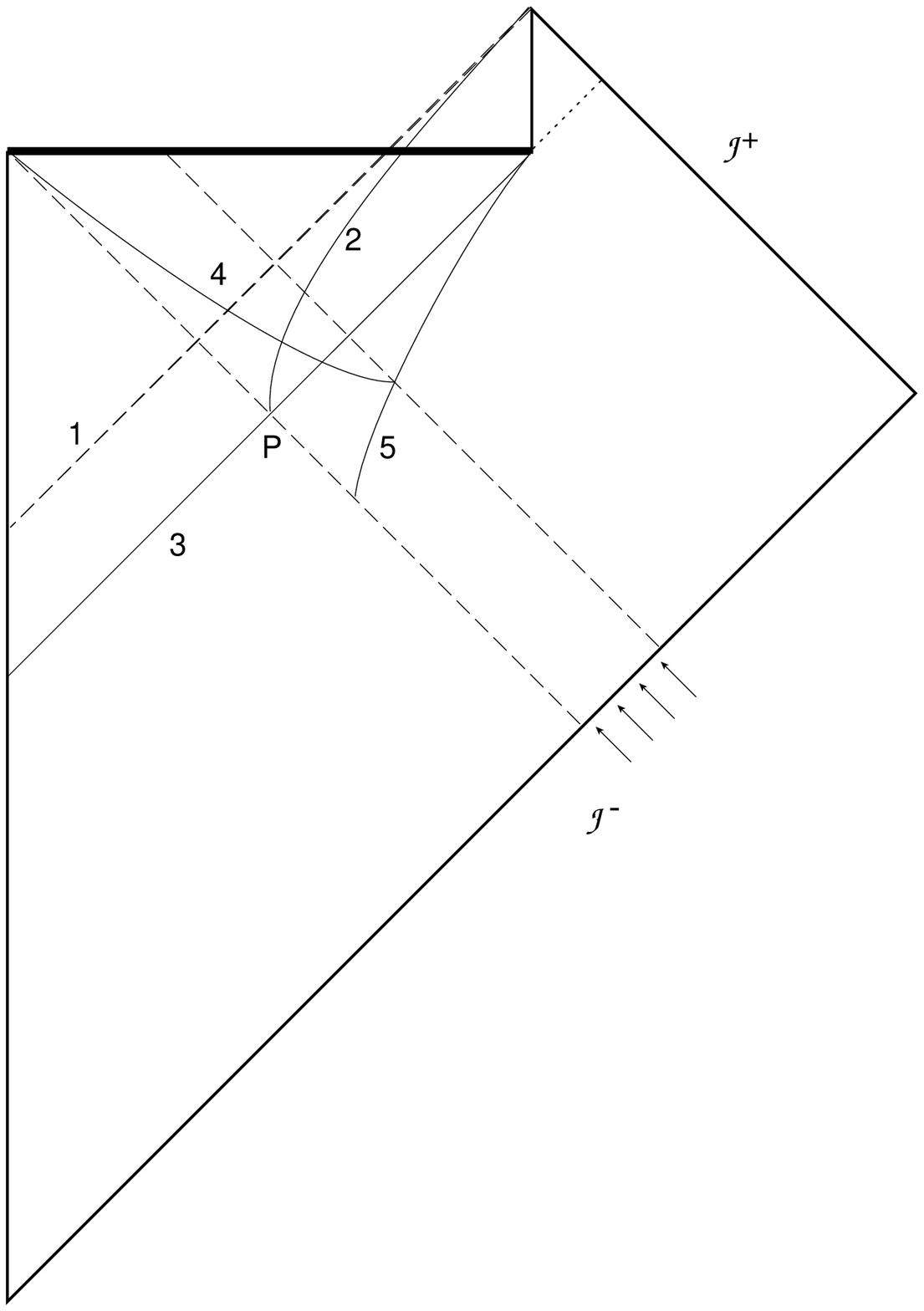}}

Consider the process of black hole formation and evaporation, and
let us assume that the final geometry is given by the Minkowski metric
with radial coordinate $r_*$ and time $t$.
We will  refer to different horizons (see fig. 1):
\smallskip
\n 1) {\it Ultimate horizon}: The null line passing through
the boundary point of the future null infinity ${\cal J}_+$. Here the
Minkowski time $t$ of the outside observers goes to infinity.

\n 2) {\it Hidden horizon}:
To construct it,
a null line is drawn from ${\cal J}_-$ to the point where the boundary
becomes space-light. This line intersects the event horizon at some point $P$.
Let  $r_{*h}$ be the value of $r_*$ at this point. The {\it hidden horizon}
is the time-like surface of space-time points with $r_*=r_{*h}$. It is
placed between the event and the ultimate horizon.
It corresponds to the minimum value that the Rindler coordinate takes
just before the disappearance of the black hole. It provides an upper
bound for the redshift  that an external
observer can detect on an outgoing Hawking particle.

\n 3) {\it Event horizon}: Defined as usual in general relativity
as the boundary of the black hole.

\n 4) {\it Apparent horizon}: Defined in the standard way
as the boundary of the total trapped surface
(see e.g. ref. \wald\  and appendix A).

\n 5)  {\it Stretched} and {\it global apparent horizons}:
Let $r_{*e}$ be the value of $r_*$ at the endpoint of
Hawking radiation (the endpoint is  the intersection between the event
horizon and the singularity line). We define the {\it stretched horizon}
to be the time-like surface of space-time points with $r_*=r_{*e}$.
The {\it global apparent horizon} is the time-like surface which
coincides with the apparent horizon after
the incoming energy flux has stopped and, for earlier times, it is
defined by analytic continuation.
For large black holes the stretched and the global apparent horizons are
the same object, as   will be shown below.
 This horizon is  of relevance in describing the evaporation
of a black hole. In particular,
the bulk  of  Hawking radiation  originates at
the global apparent horizon. As the black hole evaporates, the
radius on this horizon recedes in correspondence with the mass loss
of the black hole.
The proper distance from these horizons to the event horizon,
$\delta s=\int _{e.h.}^{s.h.} dr \sqrt{g_{rr}}$,
 and the  local Unruh temperature, are of Planck magnitude
(times a constant depending also on the number of propagating
fields).
The stretched (or global apparent) horizon coincides with the event horizon
in the limit where
$l_{\rm Planck }$, or $\h $, or the number $N$ of propagating local degrees of
freedom goes to zero.

%\n 6) {\it Gauge} (or {\it String}  ) {\it horizon}:
%This horizon will play no role in the present analysis, but
%it is a useful object for describing the behavior of infalling matter as seen
%by an external observer. In particular, it is central in the study of
%the final phase of Hawking evaporation \suss . It is the layer where the local
%temperature of the Unruh radiation is of order $gM_{\rm Planck} $ (Hagedorn
%temperature), where $g$ is a gauge (or string) coupling.
%For a particle lying on this horizon the gauge interactions
%are comparable to the gravitational interactions.

\smallskip

In the context of the information
loss problem, the ``stretched" horizon was  discussed in ref. \stu .\foot{
For the role of a {\it stretched horizon} in classical black hole physics,
see ref. \membr .}
 There
it was described as a time-like surface where the local Unruh temperature
is of Planck order, but the exact value was not specified.
The above definition determines a very specific
time-like surface.
It is  important here to determine the exact
location of the global apparent horizon for the calculation of the
Bekenstein entropy.

The present results may be interpreted as follows.
Let us suppose we use the standard field-theoretic formula
for entropy, and then ask where  the brick wall must be placed
in order to agree with the usual formula.  For
two-dimensional black holes the wall must be placed at an exponentially small
distance from the ultimate horizon, with a particular
$N$ dependence. It turns out that the same dependence governs the separation
between the ultimate horizon and the event horizon.
  It is thus natural to locate the entropy in the inaccessible
region between these two horizons.
In four dimensions, the same reasoning will locate the entropy on the
stretched horizon.
Since the bulk of the Hawking radiation is in
the region which is in causal contact with the stretched horizon,
and it is causally disconnected from
the hidden horizon, the above suggests that in two dimensions
the quantum mechanical information cannot reappear in the Hawking
radiation, whereas in four dimensions Hawking particles can
conceivably carry the information.

%%%%%%%%%%%%%%%%%%%%%%%%%%%%%%%
\newsec{Entropy in two-dimensional gravity}
%%%%%%%%%%%%%%%%%%%%%%%%%%%

The black hole entropy in two-dimensional dilaton gravity
(as can be derived, e.g. from thermodynamics or as a surface term)
is given by:
\eqn\entrp{
S= {2M\ov \l}\ \ .
}
Different discussions of black hole entropy in two-dimensional dilaton
gravity can be found e.g. in ref. \entropy .
A statistical mechanical (or fine-grained) explanation of the entropy
\entrp\ is unknown.
To derive the black hole entropy
from statistical mechanics, we will count field eigenmodes near the horizon
following refs. \thoof\ and \susugl .

Let us first briefly review the
approach where the black hole  is treated as a classical fixed background
\susugl . In Kruskal coordinates the classical black hole geometry
formed by the collapse of a shock wave at time $x^+_0$
is given by
\eqn\dos{
ds^2=-{dx^+dx^-\ov {M\ov \l} -\ll x^+(x^-+ p)  }\ ,\ \ \
e^{-2\phi }={M\ov \l} -\ll x^+(x^-+ p )\ ,\ \ \ p\equiv {M\ov\l^3 x^+_0}\ .
}
Rindler coordinates are introduced by
\eqn\tres{
{M\ov\l} \sinh ^2(\l R)=-\ll x^+(x^-+ p)\ , \ \ \ e^{2\l T}=-{x^+\ov x^-+p}\ ,
}
\eqn\cuatro{
ds^2=dR^2 - \tanh ^2 (\l R) dT^2  \ ,\ \ \
\phi=-\ha \log \bigg[ {M\ov\l}\cosh ^2 (\l R) \bigg]\ .
}
Near the event horizon, $R\cong 0$, and the black hole metric
is well approximated by the Rindler metric $ds^2=dR^2 -\ll R^2 dT^2$.
Consider $N$ massless scalar fields $f_i $ propagating in this background.
In Kruskal coordinates the field equation is simply $\dd_+\dd_-f_i=0$.
The solution in Rindler coordinates can be written as
$f(R,T)=\vp (R) e^{i\w T}$, where
$$
\vp (R)=A \cos(\w \s)+B \sin (\w\s)\ ,\ \ \l \s= \log (\l R)\ .
$$
In the spirit of the ``brick wall" model, we will introduce an ultraviolet
cutoff $R=\ep $ and, in addition, an infrared cutoff at $R=L$. Assuming
periodic boundary conditions, the eigenvalues for $\w$ are then
$\w_n ={\pi  n\ov l}$, where $l={1\ov\l} \log (L/\ep),\     n=1,2,...$.
The free energy is given by
\eqn\freef{
F(\b )={N\ov \b}\sum _{n=1}^\infty \log (1-e^{-\b\w_n })\ .
}
For large $L$,
$$
\sum _{n=1}^\infty \log (1-e^{- na })=-\sum _{m=1}^\infty {e^{-ma}\ov
m(1-e^{-ma} ) } \cong - {\pi^2\ov 6} {1\ov  a}\ ,\ \
a={\pi \l \b \ov \log (L/\ep)}\ .
$$
Thus
\eqn\freeff{
F(\b )=-{\pi N\ov 6 \l \b^2 } \log {L\ov \ep}\ .
}
Using $S=\b^2 {\dd F\ov \dd\b }$ and the fact that $\b=2\pi/\l $,
 the entropy is found to be given by
\eqn\ssss{
S={N\ov 6}\log {L\ov \ep}\ .
}
The entropy has an ultraviolet
divergence as $\ep\to 0$ due to an accumulation of modes near the horizon.

Now it  will be shown that the result is actually finite upon inclusion of
back-reaction effects.
The   evaporation makes that the event
and the ultimate horizon split.
The field-theoretical formula cannot be extrapolated up to the ultimate
horizon, because the black hole has disappeared long before.
%At most, this formula can be used  up to the minimum value that the
%Rindler coordinate takes just before the disappearance of the black hole.
As a result,   physical quantities relevant for time-like external observers
always remain finite,  taking their
maximum value near the event horizon.
In particular, the maximum redshift that an outside observer can
detect on an outgoing particle is given in terms of the
value of $g_{00}$ on the minimum value that the Rindler coordinate
takes  just before the disappearance of the black hole.
The ``hidden horizon"
regularizes all quantities which would classically diverge
if the evaporation is not included.

Let us consider an arbitrary distribution $T_{++}(x^+)$ of incoming matter
with an energy density flux above the threshold for black hole formation
\rst . The collapse starts at time $x^+_0$ and finishes at time $x^+_1$.
In Kruskal coordinates, the resulting time-dependent geometry is given by
$ds^2=-e^{2\rho} dx^+dx^-,\ \rho=\p $ and
\eqn\metrica{
2\k\p +e^{-2\p }=-\ll x^+\big( x^-+ \l^{-2} P_+(x^+) \big)
+{1\ov \l} M(x^+)-\k \log(-\ll x^+x^-)\  ,
}
$$
\k={N\ov 48}  \ ,\ \ P_+(x^+)=\int_{x_0^+}^{x^+} dx^+ T_{++}(x^+)\ ,\ \
M_+(x^+)=\l \int_{x_0^+}^{x^+} dx^+ x^+ T_{++}(x^+)\ .
$$
The curvature singularity is at $\p (x^+,x^-)=-{1\ov 2}\log \k $.
By definition the apparent horizon is at $\dd_+ \p=0$ \rstf\
(see   appendix A).
This is the time-like curve
\eqn\appar{
\ll x^+(x^-+ \l^{-2} P_+(x^+))=-\k\ .
}
After the incoming flux has stopped, $x^+>x^+_1$, we have
 $P_+(x^+)=P_+(x^+_1)\equiv  \ll p $ and
$M_+(x^+)=M_+(x^+_1)\equiv  M $,  and the equation of the
apparent horizon becomes simply $\ll x^+(x^-+p)=-\k$.
The endpoint of the black hole, i.e. the point where the singularity
becomes light-like, is at the intersection of the apparent horizon and
the singularity line $\p (x^+,x^-)=-{1\ov 2}\log \k $. This occurs
at
\eqn\endpo{
x^+_e={\k\ov \ll p} (e^{M/\k\l}-1)\ ,\ \ \
x^-_e=- p (1-e^{-M/\k\l})^{-1}\ .\ \ \
}
The event horizon is thus the line $x^-=x^-_e,\ x^+ < x^+_e $.
At $x^+>x^+_e\ , x^-=x^-_e $ the geometry is matched with the Minkowski
vacuum, $ds^2=-d\tau^2 +d\s^2 ,\ \p= -\l\s $, where
\eqn\torto{
e^{2\l\s}=-\ll x^+(x^-+p)\ ,\ \ e^{2\l\tau }=-{x^+\ov x^-+p}\ .
}
The ultimate horizon is at $\tau=\infty $, i.e. the null
line $x^-=-p$. To determine the stretched horizon and the
hidden horizon, we apply the definitions given in the previous section.
Now $r_*$ corresponds to $\s $. The stretched horizon is at
$\s=\s_e={1\ov 2\l}\log [ -\ll x^+_e(x^-_e+p)]$, which is the
hyperbola
\eqn\stretch{
\ll x^+(x^-+ p)=-\k \ .
}
As anticipated in section 1, this coincides with the apparent horizon for
$x^+>x^+_1$, and it is in this sense an analytic
continuation of the apparent horizon in the region $x^+<x^+_1$.

For large black holes the location of the global apparent horizon can
also be determined
without having an exact solution of the back-reaction problem. For $x^+>x^+_1$
the equation
$\dd_+ \p=0$ can be written as
\eqn\apotr{
0={d e^{-2\p}\ov dx^-}-{\dd e^{-2\p}\ov \dd x^-}\cong
{d e^{-2\p}\ov dx^-}+ \ll x^+ \ .
}
Since the value of $e^{-2\p}$ at the horizon is the mass of the black hole,
$e^{-2\p _h}=M/\l $, we can compute ${d e^{-2\p}\ov dx^-}$ from the
mass loss rate. This can be obtained by differentiating the
Bondi mass \rst ,
$${dM(x^-)\ov dx^-} \cong {\k \l \ov x^-+p}\ . $$
Inserting this into eq. \apotr\ the global apparent horizon is
found at $\ll x^+(x^-+p)=-\k $, in exact
agreement with eq. \stretch .

Similarly, using $x^+_0(x^-_e+ p)= -px^+_0 (e^{M/\k\l}-1)^{-1}$,
 the hidden horizon is found at
\eqn\hidd{
x^+(x^-+ p)= -px^+_0 (e^{M/\k\l}-1)^{-1}\ .
}

Now we would like to count all field configurations that
are outside the black hole.
%This corresponds to the shaded region in fig. 3.
%\bigskip
%Fig. 3: The shaded area of this Penrose diagram is the region
%of relevance for an external observer (outside the event horizon
%\bigskip
The main contribution comes, as calculated above, from field configurations
near the event horizon, where the geometry (for a large black hole)
is very accurately described by the Rindler metric.
The previous calculation
applies, except that now the Rindler coordinate,
defined by
$MR^2=-\l x^+(x^-+p)$, does not start at zero. The minimum value that the
Rindler coordinate takes before the black hole disappears
is at the intersection of the event horizon and the hidden horizon:
\eqn\rinmin{
R_{\rm min}^2={\l  px^+_0\ov M} (e^{M/\k\l}-1)^{-1}\ .
}

The curve $R=R_{\rm min}$ is  the hidden horizon given
by eq. \hidd .
 Thus we find that the fine-grained entropy associated
with fields near the event horizon is
\eqn\diez{
S={N\ov 12}\log {L^2\ov R_{\rm min}^2}=
{4M\ov \l} +{N\ov 12}\log \bigg( {L^2M\ov \l p x^+_0 } \bigg)
+O(e^{-M/\k\l })\ .
}
The first term represents the black hole entropy in 2d dilaton gravity.
The second term could be regarded as an additional quantum correction,
but here it is of no relevance.\foot{ Note that
for a shock wave collapse
this term simplifies,
$\log \big({ L^2M\ov \l p x^+_0} \big) =2 \log (\l L) \ $. }
The interesting
point here is having obtained $S={4M\ov \l}$  from just statistical
mechanics and quantum field theory. It is finite, and it differs
from eq. \entrp\  by only a factor of 2 (this may be due to different reasons,
 we will not speculate on this here).
 Interestingly, the factor $N$ has cancelled out and
$S$ is of order $1/\h $ once
$\h $ is restored in the formulas.

By separating right and left moving-mode contributions,
 eq. \ssss\ can be written in the following way:
\eqn\ssttt{
S={N\pi T\ov 6\l }\bigg[\log {x^-_L+p\ov x^-_\ep+p}+\log {x^+_L\ov
x^+_\ep}\bigg]
\ .}
Let us consider the entropy density, ${\cal S}={\dd S\ov\dd \s^-}\ ,\
\s^-=\tau-\s $ (see eq. \torto ). Using the thermodynamic relation, $T d{\cal
S}=d{\cal E}$,
 the formula for the outgoing energy density is recovered:
${\cal E}={N\pi T^2\ov 12}=N\int_0^\infty {d\w\ov 2\pi} {\w\ov e^{\w/T}-1}$.
For $x^->x^-_e$ there is no more Hawking radiation, so in this region one has
 ${\cal E}=0$. Thus the identification   $x^-_\ep=x^-_e$, that was made in
deriving eq. \diez , is nothing but the statement that the Hawking
energy-density flux stops at $x^-=x^-_e$.
The correct dependence on $M$  has emerged in eq. \diez ,
  by just taking into account the physical fact that after some finite
time the black hole disappears into the vacuum.

%%%%%%%%%%%%%%%%%%%%%%%%%%%%%%%
\newsec{Entropy in four-dimensional gravity}
%%%%%%%%%%%%%%%%%%%%%%%%%%%
The calculation of the entropy
in the case of the Schwarzschild black hole
using the brick wall model is similar to the calculation in the first part of
the previous section.  The Rindler coordinate is
\eqn\rindd{
R^2=8GM(r-2GM)\ .
}
The following expression is found in terms of a Rindler  cutoff $R=\ep $
\susugl \ (see also  ref. \thoof ):
\eqn\oncez{
S={N A \ov 360 \pi \ep^2}\ ,\ \ \ A=16\pi G^2M^2\ .
}

 Let us now take into account the back reaction in the geometry.
Although  a full treatment including the evaporation cannot be performed
by using exact analytic methods, for large black holes it is possible to
make some very accurate estimates.
The minimum value that  the Rindler coordinate $R$ can take outside
the event horizon occurs at the point where the hidden horizon begins.
This value is calculated in   appendix B,
\eqn\minii{
R_{\rm min}^2=\ep ^2= {N G\ov 60\pi } e^{-bGM^2}\ ,\ \ \ \ b={320\pi\ov N}\ .
}
This gives
\eqn\once{
S= {A\ov 6G }  e^{b GM^2}\ .
}
Equation \once\ represents the entanglement entropy  as predicted by
quantum field theory;
it is finite, but  exponentially
larger than the Bekenstein entropy, eq. \uno .
Thus, unlike the two-dimensional case, the straightforward calculation of the
entropy that takes into account the evaporation does not give a reasonable
result. This can be viewed as a failure of conventional quantum field theory
to correctly describe physics in the vicinity
of the event horizon \thoof . A short-distance cutoff is needed before.

{}From the point of view of the information problem, this is actually
fortunate:
if the entropy was to be associated with degrees of freedom near the
hidden horizon (or near any region at an exponentially small distance
from the event horizon), this would strongly suggest that the information
could not possibly come out with  the Hawking radiation.
Hawking radiation originates in the region causally connected
with the global apparent horizon. Let us thus postulate --following
  \refs{ \thoof, \stu } -- that the quantum
mechanical information reappears in the Hawking radiation.
The natural place for the wall in this scenario  is
the global apparent horizon. This is a very specific surface,
so this ansatz will provide an unambiguous  prediction for the Bekenstein
entropy. If this prediction does not agree
with the expression for the black hole entropy, eq. \uno, then this
will mean that this ansatz cannot be correct.
  Surprisingly, we will find an almost exact agreement.

Let the final Minkowski metric be given by
\eqn\minka{
ds^2=-dt^2+dr_*^2+r_*^2 d\Omega^2 \ .
}
Let us introduce standard Kruskal coordinates
\eqn\kkk{
-{1\ov 2MG} V(U+2MG)=2MG e^{r_*/2MG}\ ,\ \ \ -{V\ov U+2MG}=e^{t/2MG}\ .
}
or (cf. eq. \torto )
\eqn\kkkk{
V=2MG e^{v/4MG}\ ,\ \ \ U+2MG=-2MG e^{-u/4MG}\ ,\ \ \ \ v,u=t\pm r_*\ .
}
%On the plane $U-V$ the global apparent horizon is the time-like curve
%given by the equation $r_*(U,V)=r_{*e}$, where $r_{*e}$ is the value of $r_*$
%at the endpoint of the evaporation.
After the incoming flux of collapsing matter stops,
apparent and global apparent horizons coincide.
For spherically symmetric configurations the
apparent horizon is at ${\dd r(U,V)\ov \dd V}=0$ (see appendix A).
For a large black hole (away from the endpoint) the  geometry
is given by the Schwarzschild metric  with
\eqn\trece{
2MG (r-2MG)e^{r\ov 2MG}=-V(U+2MG)\ .
}
Near the horizon, $r\cong 2MG$, and we can solve
this transcendental equation for $r$
iteratively. We are only interested in the leading part, which is
\eqn\itera{
r \cong -{1\ov 2eMG} V(U+2MG) +2MG +O\bigg( {M_{\rm Planck}^2\ov M^3}\bigg)\ .
}
We will determine the global apparent horizon by solving the equation
${\dd r(U,V)\ov \dd V}=0$ after the incoming flux of collapsing matter has
stopped, just as we did in the previous section
(see eq. \apotr ).
This equation can be written as
\eqn\aapp{
0={d r_{\rm GAH}\ov d U}-{\dd r_{\rm GAH}\ov \dd U}\cong
{d r_{\rm GAH}\ov d U}+{1\ov 2eMG} V
%\bigg[1 + O\bigg( {M^2_{\rm Planck}\ov M^2}\bigg)\bigg]
\ .
}
For a large Schwarzschild black hole, $r_{\rm GAH}\cong 2MG$, so that
\eqn\masss{
{dr_{\rm GAH}\ov dU} \cong 2G {dM\ov dU}\ .
}
Now, in the
vicinity of the horizon, a black hole loses mass at a rate
\eqn\docec{
{d M\ov d  u}=-N{\pi^2\ov 15} T^4 A= -{N\ov 3840 \pi G^2} {1\ov M^2}\ .
}
Thus
\eqn\doce{
{d M\ov d  U}={N\ov 960 \pi G  (U+2MG)} {1\ov M}\ .
}
Using eqs. \aapp , \masss \ and \doce , we find the following equation
for the (global) apparent horizon (cf. eq. \stretch ) :
\eqn\gah{
V(U+2MG)\cong -{NeG\ov 240\pi } \ .
}
It is clear that for a large black hole (i.e. $M>>M_{\rm Planck}$)
the global apparent horizon coincides with the stretched horizon. Indeed,
the apparent horizon must pass through the black hole endpoint (from the
definition of ``endpoint"). Comparing with eq. \kkk , we see
that the equation of the global apparent horizon \gah\ (which is correct
up to terms of $O(M^2_{\rm Planck}/M^2)$) can be simply written as
$r_*=r_{*e}$. This is precisely our definition of stretched
horizon.

In terms of $r$ this is the time-like surface  $r=r_{\rm S.H.}=2MG+\delta $ ,
with (see eq. \trece )
\eqn\gahr{
r_{\rm S.H.}\cong 2MG + {N\ov 480\pi }{1\ov M}\ .
}
Let us check that the proper distance $d s$ from this surface to the
event horizon is indeed of Planck order. We have  ($\delta<< 2MG$):
\eqn\prope{
d s= \int_{2MG}^{2MG+\delta }
{dr\ov \sqrt{1-{2MG\ov r}} }\cong 2\sqrt{ 2MG\delta }=
\ha \sqrt{ {NG\ov 15\pi } }\ .
}
We see that the proper distance increases with the square root of the number
of scalar fields.
Using the expression for the  Rindler coordinate,  eq. \rindd ,
we find $R_{\rm  S.H.}^2= {NG\ov 60\pi}$, whence we finally obtain
\eqn\sfinal{
S={2\ov 3} {A\ov 4 G \h }\ ,
}
which differs from the Bekenstein entropy   by only a factor of 2/3.
\smallskip

To summarize, in two dimensions
the standard field-theoretic formula for the entropy,
excluding contributions beyond $R_{\rm min}$,  reproduces the usual
expression for the black hole entropy up to a numerical constant of order 1.
In four dimensions, quantum field theory predicts   too large a
fine-grained entropy given by eq. \once .
An ultraviolet cutoff is necessary  before the event horizon,
perhaps due to a breakdown of conventional quantum field theory \thoof.
For the information to come out in the Hawking radiation,
it is natural to locate the entropy
on the global apparent horizon, since this is precisely the place
where most Hawking particles originate.
The location of this horizon was calculated and,
strikingly,  the  Bekenstein entropy formula has  emerged,
independent of the number of
propagating fields, with the correct dependence on the   mass, the Newton
constant  and $\h $.
These  results may  be viewed as an indication that in four dimensions
the quantum mechanical information
is encoded in long-time correlations in the Hawking radiation
\refs{ \thoof, \stu , \svv }.
\bigskip
%%%%%%%%%%%%%%%%%%%%%%%%%%%%%%%%%%%%%%%%%%%%%%%%%%%%%%%%%%%%%%%%%%%%
\noindent $\underline{\rm Acknowledgements}$:
%%%%%%%%%%%%%%%%%%%%%%%%%%%%%%%%%%%%%%%%%%%%%
\noindent The author would like to thank L. Susskind and E. Verlinde for
useful discussions.

\appendix {A}{Apparent horizon for spherically symmetric configurations}

In this appendix   we discuss the connection between the two-dimensional
definition of apparent horizon,
$\dd_+\p\dd_-\p=0$, and the general four-dimensional definition (a further
discussion on the latter can be found, e.g. in ref. \wald ).

Let $C$ be a three-dimensional manifold with boundary $S$.
Let $\xi _\mu$, $\mu=0,1,2,3$, be the vector field of tangents to a
congruence of
outgoing null geodesics orthogonal to $S$. $C$ is a {\it trapped region}
if the expansion $\theta=\nabla_\mu\xi^\mu $ is everywhere non-positive
on $S$, $\theta \leq 0$. The {\it apparent horizon} ${\cal A}$ is the
boundary of the {\it total trapped region}, the latter defined as the closure
of the union of all trapped regions. A corollary of this definition
is that $\theta=0$ on ${\cal A}$.

  Let us now contemplate metrics of the form
 \eqn\bmet{
ds^2=g_{ij}(x^0,x^1) dx^idx^j+ \exp [-2\phi (x^0,x^1)] d\Omega ^2
\ , \ \ \ \ i,j=0,1\ .
 }
 In this spherically symmetric space-time, we have
$\xi_\mu=\{ \xi_0,\xi_1,0,0 \}$, and the geodesic equation reduces to
\eqn\geob{
\xi^i\nabla_i\xi^j =0\ \ ,
}
i.e. the two-dimensional geodesic equation. Since in this dimensionally
reduced configuration there is only one family of outgoing
null geodesics, a trapped region is the total trapped region, and the
condition determining the apparent horizon simply becomes
\eqn\bthet{
\theta=0\ .
}
{}From eq. (A.1) one easily obtains
\eqn\bthett{
\theta =\theta ^{(2)} -2\xi^j\dd_j\phi\ \ ,
}
where $\theta ^{(2)} \equiv \dd_i\xi^i+\Gamma_{ij}^i\xi^j\ $.
 Let us denote $B_{ik}=\nabla _i\xi_k $. By using the geodesic equation,
$\xi^iB_{ik}=0$,
and from the fact that $\xi $ is null, $\xi^i\xi_i=0$,   the following
relations can be derived:
\eqn\buub{
\xi^1B_{11}=-\xi^2B_{21}\ ,\ \ \ \xi^1B_{21}=-\xi^2B_{22}\ ,\ \ \ B_{12}=B_{21}
\ \ ,
}
from which we obtain
\eqn\tttb{
\eqalign{
\theta^{(2)} &=g^{ij}B_{ij}=B_{11}\big[ g^{11}-2{\xi^1\over\xi^2}g^{12}
+\big({\xi^1\over\xi^2}\big) ^2g^{22}\big]  \cr
&=0 \ ,\cr}
}
where we have used $\xi^1\xi_1=-\xi^2\xi_2$.   We thus see that the
two-dimensional expansion parameter is identically zero (in particular,
this means
that an intrinsically two-dimensional apparent horizon cannot be defined).
Now, by using eqs. (A.3), (A.4) and  \tttb , we find that the
condition defining the
apparent horizon becomes
\eqn\apphb{
\xi^i\dd_i\phi = 0\ .
}
Since $\xi $ is null, eq. \apphb\ implies $\xi_i = f(x) \dd_i\phi $, where
$f(x)$ is a scalar function. Therefore the condition (A.3) translates to
\eqn\gggb{
g^{ij}\dd_i\phi\dd_j\phi=0\ ,
}
or, in the conformal gauge, $ \dd_+\phi\dd_-\phi=0 $ ,
 which thus determines the location of  the apparent horizon ${\cal A}$.

\appendix {B}{Hidden horizon in four dimensions}

Here we calculate the value of the
Kruskal coordinates at the endpoint of an evaporating four-dimensional
black hole, as well as
the location of the hidden horizon.
Let us assume that the black hole is formed by the collapse of a shock wave at
advanced time $v=v_0=0$ (so that $V_0=2MG$).
The black hole will Hawking-radiate
at a rate
\eqn\auno{
{d M(v)\ov d  v}=-{N\ov 3840 \pi G^2} {1\ov M^2(v)}\ ,
}
i.e.
\eqn\ados{
{N\ov 1280 \pi G^2}v=M^3-M^3(v)\ .
}
Thus  (see eq. \kkkk )
\eqn\atres{
V_e=2MG e^{bGM^2}\ ,\
\ \ \ b={320\pi\ov N}\ .
}
Since the apparent horizon passes through the endpoint,
we can use eq. \gah\ to determine $U_e$:
\eqn\acuatro{
U_e+2GM=-{Ne\ov 480 \pi M}e^{-bGM^2}\ .
}
By definition the hidden horizon is at $r=r_{*h}$,
where
\eqn\hidhor{
2MG e^{r_{*h}/2MG}=-{1\ov 2MG} V_0(U_e+2MG)\ .
}
This is the hyperbola
\eqn\hidhor{
 V (U +2MG)=-{NeG \ov 240\pi} e^{-bGM^2}\ .
}
The Rindler coordinate is $R^2=8MG(r-2MG)\cong -4e^{-1} V(U+2MG)$.
 This gives eq. \minii .

\vfill\eject
  \listrefs
\vfill\eject
\end